\documentclass[12pt]{article}
\usepackage{amsmath}
\usepackage{graphicx}
\begin{document}
\baselineskip=18 pt
\begin{center}
{\large{\bf Causality violation in plane wave spacetimes}}
\end{center}

\vspace{.5cm}

\begin{center}
{\bf D. Sarma}\footnote{debojitsarma@yahoo.com}, 
{\bf M. Patgiri}\footnote{mahadev@scientist.com} and
{\bf F. Ahmed}\footnote{faiz4U.enter@rediffmail.com} \\
{\it Department of Physics, Cotton College,}\\
{\it Guwahati-781001, India}
\end{center}

\vspace{.5cm}

\begin{abstract}
It is widely held that plane wave spacetimes are causally well behaved which means that they are free from causal pathologies such as closed timelike curves or closed null geodesics. In this paper, we show that one can get closed null geodesics in a plane wave metric by analytically solving the geodesic equations.
\end{abstract}



\vspace{.5cm}

\section{Introduction}

One of the most fascinating aspects of the general theory of relativity is the fact that certain solutions of Einstein's field equations admit closed paths which may be followed by material particles or by light (or other massless particles) effectively allowing them to travel to their past. Such solutions have been with us, and with no resolution of the problem, since G\"odel surprised Einstein with a solution of the field equations with closed timelike curves (CTCs) \cite{Godel}. Since then, a number of spacetimes with closed causal curves (CCCs), which include closed timelike curves (CTCs), closed null geodesics (CNGs) and closed timelike geodesics (CTGs), have been found.  
Some examples of spacetimes with CTCs are the van Stockum spacetime \cite{Sto}, where CTCs were shown to exist in \cite{Tip}. Bonnor and Steadman \cite{Bon} have shown that a spacetime with two spinning particles, under special circumstances, allow for the existence of CTGs. Another well-known example is Gott's \cite{Gott} spacetime with two moving cosmic strings. All the examples mentioned above are solutions of the field equations involving various distributions of matter. There are also vacuum spacetimes with closed causal curves. For instance, the NUT-Taub metric \cite{Taub} and the time machine spacetimes found by Ori \cite{Ori1,Ori2} are examples of spacetimes satisfying the vacuum Einstein equations and admitting CTCs. The examples cited above by no means exhaust the list of spacetimes which admit CCCs. 

Plane wave spacetimes, which are a subset of {\it pp} waves, form one of the most interesting class of solutions of Einstein's field equations for a variety of reasons and have been an intense field of study for string theorists. For one, they are exact solutions in string theory for all orders of perturbation of string tension \cite{Amat,Horo}. Additionally, they provide maximally supersymmetric background for propagation of strings \cite{Blau}. Another point of interest is the Penrose limit \cite{Penr}, which would mean that in the neighbourhood of a null geodesic, any spacetime resembles a plane wave. Plane wave spacetimes are not globally hyperbolic \cite{Penro}, but is well accepted that they are causally stable \cite{Penro,Hawk,Ger} {\it i.e.}, they are free from CTCs and CNGs.

In the context of explicit connection between plane waves and CTCs, it has been shown that maximally supersymmetric plane waves \cite{Blau} is related to the supersymmetric G\"odel universe \cite{Gaun} via T-duality \cite{Boyd,Harm}. Hubeny {\it et. al.} have studied CTCs in plane waves \cite{Hub1} and have concluded that CTCs are introduced by quotienting instead of by T-duality. 
They also demonstrate that plane waves are stably causal thus implying that this class of spacetimes does not admit CCCs. In certain compactified plane wave spacetimes, CTCs have been found with CTGs pushed out infinity \cite{Bre}. The techniques  outlined above, which introduce causal pathologies of this nature in plane wave spacetimes, have been termed special procedures in this paper. It appears, however, that one can find plane wave metrics which admit CCCs without employing special procedures. Ori's famous time machine spacetime \cite {Ori2}, which admits CTCs, is indeed a vacuum plane wave as has been stated in his paper. In this paper, we will also write down a vacuum, plane wave spacetime and by analytically solving the geodesic equations demonstrate that this spacetime admits CNGs.

The plan of the paper is as follows: in section {\it 2},  we present a metric for which we analytically solve the geodesic equations and show that a particular set of solutions of these equations represent CNGs. In section {\it 3}, the spacetime is analyzed in detail to show that the spacetime belongs to the class of plane waves. Section {\it 4} is the concluding one.

\section{A spacetime with CNGs}

Consider the cylindrically symmetric metric
\begin{equation}
ds^2=dr^2+2{\sqrt 2}\,z\,dr\,d\phi+r^2\,d\phi^2+dz^2+dw\,d\phi
\label{1}
\end{equation}
where $\phi$ is a periodic coordinate $\phi\sim\phi+2\pi$. We have used coordinates $x^1=r$, 
$x^2=\phi$, $x^3=z$ and $x^4=w$. The ranges of the other coordinates are $-\infty < z < \infty$, $0 < r < \infty$ and $-\infty < w < \infty$. The metric has signature $(+,+,+,-)$ and the determinant of the corresponding metric tensor $g_{\mu\nu}$, $det\;g=-\frac{1}{4}$.
The line element (\ref{1}) defines a Ricci flat spacetimes
having no curvature singularities. The metric admits a covariantly constant null vector which indicates that it is a {\it pp} wave. In fact, as we shall show subsequently, these are plane wave spacetimes which are a sub-class of {\it pp} waves.

The geodesic equations are
\begin{equation}
\ddot{x}^{\mu}+\Gamma^{\mu}_{\alpha\beta}\dot{x}^{\alpha}\dot{x}^{\beta}=0
\label{n9}
\end{equation}
where the dots denote differentiation with respect to an affine parameter $\lambda$ along the geodesic.
Writing out the equations (\ref{n9}) in full, we have 
\begin{eqnarray}
\label{21}
\ddot{r}&=&\dot{\phi}\left (r\,\dot{\phi}-\sqrt{2}\,\dot{z}\right ) \quad ,\\
\label{22}
\ddot{\phi}&=& 0 \quad ,\\
\label{23} 
\ddot{z}&=& \sqrt{2}\,\dot{r}\,\dot{\phi} \quad ,\\
\label{24}
\ddot{w}&=&-2\,r\left (2\,\dot{r}+\sqrt{2}\, z\,\dot{\phi}\right )\dot{\phi}-2\left (\sqrt{2}
\,\dot{r}-2\, z \,\dot{\phi}\right )\dot{z} \quad .
\end{eqnarray}

A set of solutions for the coupled differential equations (\ref{21}), (\ref{22}), (\ref{23}) and 
(\ref{24}) are 
\begin{eqnarray}
\label{25}
r&=&\frac{1}{a}\left [A_1\,\sin(a\,\lambda)-A_2\,\cos(a\,\lambda)\right ] \quad ,\\
\label{26}
\phi&=&a\,\lambda+b \quad ,\\
\label{27}
z&=&A_3-\frac{\sqrt{2}}{a}\left [A_1\,\cos(a\,\lambda)+A_2\,\sin(a\,\lambda)\right ]\quad , \\
\label{28}
w&=&\frac{A_4}{2}+2\sqrt{2}\left(A_2\,A_3\right ) \frac{\cos (a\,\lambda)}{a}-2\sqrt{2}\left(A_1\,A_3\right)\frac{\sin(a\,\lambda)}{a} \nonumber \\
&&+\frac{3({A_1}^2-{A_2}^2)}{2\,a^2}\sin(2\,a\,\lambda)-3\,\frac{A_1\,A_2}{a^2}\cos(2\,a\,\lambda)
\end{eqnarray}
where $a, b, A_1\ldots A_4$ are arbitrary constants. It can be verified that (\ref{25}), (\ref{26}), (\ref{27}) and (\ref{28}) satisfy the differential equations (\ref{21}), (\ref{22}), (\ref{23}) and (\ref{24}) by direct substitution. 

\begin{figure}
\centering
\includegraphics{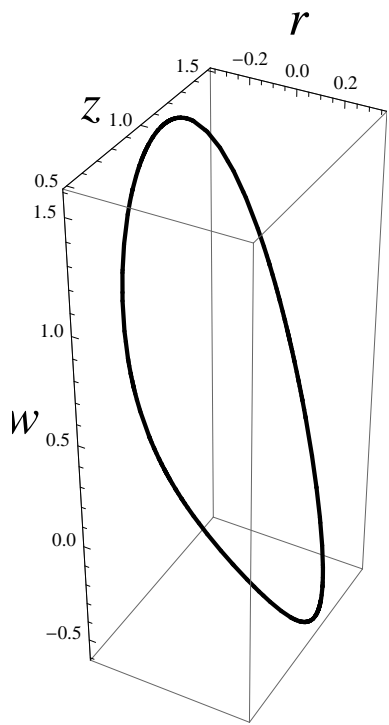}
\caption{Closed Null Geodesic}
\end{figure}

The solutions (\ref{25}), (\ref{26}), (\ref{27}) and (\ref{28}), represent parametric equations for  null geodesics in the spacetime (\ref{1}) since the tangent vector to these curves have vanishing norm, {\it i.e.} 
\begin{equation}
g_{\mu\nu}\frac{dx^{\mu}}{d\lambda}\frac{dx^{\nu}}{d\lambda}=0 \quad .
\label{null}
\end{equation}
(\ref{null}) holds for arbitrary values of the constants  $a, b, A_1\ldots A_4$. These null geodesics are also closed. We draw a parametric curve for the coordinates $r(\lambda), z(\lambda)$ and 
$w(\lambda)$ where we choose the following values of the constants for purpose of generating the graph.
\begin{equation}
a=2\,\pi, \quad A_1=2 \quad\mbox{and} \quad b=A_2=A_3=A_4=1\quad.
\end{equation}
We find that the null geodesics close for the range $0\leq s \leq 1$ and are therefore CNGs.  

\section{Analysis of the spacetime}

To establish that the metric (\ref{1}) is cylindrically symmetric, consider the Killing vector
$\boldsymbol{\eta}=\partial_{\phi}$ which has the normal form 
\begin{equation}
\eta^{\mu}=\left (0,1,0,0\right )\quad .
\label{kill1}
\end{equation}
Its co-vector is 
\begin{equation}
\eta_{\mu}=(\sqrt{2}\,z,r^2,0,\tfrac{1}{2})\quad .
\end{equation}  
(\ref{kill1}) satisfies the Killing equation $\eta_{\mu ;\nu}+\eta_{\nu;\mu}=0$. To admit an interpretation in terms of cylindrical symmetry, there should be an axis where 
$\eta_{\mu}\eta^{\mu}$ vanishes and the metric should be regular on this axis \cite{Steph}.
In our case, we find that the inner product
\begin{equation}
\eta_{\mu}\eta^{\mu}=r^2
\end{equation}
vanishes on the axis {\it i.e.} at $r=0$. Moreover, the metric is regular on the axis 
since the condition for regularity, namely, that the norm of the Killing vector $\eta^\mu$ is proportional to the square of the distance from the axis, is satisfied. Hence, the criterion for interpretation in terms of cylindrical symmetry, defined in \cite{Steph}, is satisfied.

We first show that the spacetimes represented by (\ref{1}) are of type N in the Petrov classification scheme and that they are {\it pp} wave spacetimes. We can construct the following set of null tetrads for the metric 
(\ref{1}). They are
\begin{equation}
l_{\mu}=\left (0,1,0,0\right ) \quad ,
\label{n1}
\end{equation}
\begin{equation}
n_{\mu}=\left (-\sqrt{2}\,z,-\tfrac{1}{2}r^2,0,-\tfrac{1}{2}\right ) \quad , 
\label{n2}
\end{equation}
\begin{equation}
m_{\mu}=\left(\tfrac{1}{\sqrt{2}},0,\tfrac{i}{\sqrt{2}},0\right ) \quad ,
\label{n3}
\end{equation}
\begin{equation}
\bar{m}_{\mu}=\left(\tfrac{1}{\sqrt{2}},0,-\tfrac{i}{\sqrt{2}},0\right ) 
\label{n4}
\end{equation}
where $i=\sqrt{-1}$.
The set of null tetrads above are such that the metric tensor for the line element (\ref{1}) can be expressed as
\begin{equation}
g_{\mu \nu}=-l_{\mu}n_{\nu}-n_{\mu}l_{\nu}+m_{\mu}\bar{m}_{\nu}+\bar{m}_{\mu}m_{\nu} \quad .
\label{n5}
\end{equation}
The vectors (\ref{n1}), (\ref{n2}), (\ref{n3}) and (\ref{n4}) are null vectors and are orthogonal 
except for $l_{\mu}n^{\mu}=-1$ and $m_{\mu}{\bar m}^{\mu}=1$. 

Using the set of null tetrads above, we find that, of the five Weyl scalars only
\begin{equation}
\Psi_4=-C_{\mu\nu\rho\sigma}{\bar m}^{\mu}n^{\nu}{\bar m}^{\rho}n^{\sigma}=\frac{1}{2}
\label{n6}
\end{equation}
is nonvanishing while $\Psi_0=\Psi_1=\Psi_2=\Psi_3=0$. The metric is clearly of type N
in the Petrov classification scheme. We also observe that the spacetime admits a covariantly 
constant null vector field $l_{\mu}$, {\it i.e.}
\begin{equation}
l_{\mu ;\nu}=0 \quad .
\label{n7}
\end{equation}
Such spacetimes are the plane-fronted gravitational waves with parallel rays \cite{Brink} or {\it pp} 
wave spacetimes. The metric is free from curvature singularities. The curvature invariants such as 
$R^{\mu\nu\rho\sigma}R_{\mu\nu\rho\sigma}$ and 
$R_{\mu\nu\rho\sigma}R^{\rho\sigma\lambda\tau}R_{\lambda\tau}^{\;\;\;\;\;\mu\nu}$ are non-singular, 
being constants equal to zero. This is a characteristic of {\it pp} wave spacetimes. 

Additionally, one can show that the spacetime defined by (\ref{1}) can be further classified to fall in the category of plane wave spacetimes, which is a sub-category of {\it pp} waves. Consider the transformation
$w\rightarrow 2\,v-\sqrt{2}\,r\,z$ followed by
\begin{equation}
r\rightarrow x\,\cos\frac{\phi}{\sqrt{2}}-y\,\sin\frac{\phi}{\sqrt{2}} \quad ,\quad
z\rightarrow x\,\sin\frac{\phi}{\sqrt{2}}+y\,\cos\frac{\phi}{\sqrt{2}}
\end{equation}
Then (\ref{1}) acquires the form 
\begin{eqnarray}
ds^2&=&dx^2+dy^2+2\,d\phi\,dv+\frac{1}{2}d\phi^2\left[\left(x^2-y^2\right )\cos(\sqrt{2}\,\phi)\right .\nonumber \\
&&\left .-2\,x\,y\,\sin(\sqrt{2}\,\phi)\right ] \quad.
\label{brin}
\end{eqnarray}
This is the metric reduced to the Brinkmann form. Since the metric function $g_{\phi\phi}$ in 
(\ref{brin}) is at most quadratic in $x$ and $y$, we have a plane wave spacetime.

\section{Conclusion} 

We have presented a vacuum, plane wave spacetime (\ref{1}) containing 
CNGs which contradicts the contention that plane wave spacetimes are stably causal. We have obtained these curves by explicitly solving the geodesic equations. However, a survey of the literature reveals that Ori's spacetime with CTCs \cite{Ori2} also falls in the category of plane wave spacetimes. Our spacetime differs from the one in \cite{Ori2} as the latter is a true time machine spacetime where CTCs appear after a specific instant. On the other hand, the spacetime studied in this paper belong to the category of the so-called ``eternal'' time machines among which the most famous is the one by G\"odel. In metrics with CCCs, the usual procedure for obtaining them has been to consider a subspace with all coordinates, except the periodic one, as constant and then for certain values of the radial coordinate these causality violating curves occur. For the metric studied here, direct integration of the geodesic equations is the technique employed. The spacetime represented by (\ref{1}) has not been investigated for CTCs or CTGs.

\end{document}